\documentclass[pra,aps,showpacs,twocolumn,nofootinbib,superscriptaddress]{revtex4-1}
\usepackage{amsmath}
\usepackage{amsfonts}
\usepackage{amssymb}
\usepackage{revsymb}
\usepackage{graphicx}

\def\ket#1{|\,#1 \,\rangle}

\begin{document}
\title{Branching fractions for $P_{3/2}$ decays in Ba$^+$}
\author{Zhiqiang Zhang}
\affiliation{Centre for Quantum Technologies, National University of Singapore, 3 Science Drive 2, 117543 Singapore}
\author{K. J. Arnold}
\affiliation{Centre for Quantum Technologies, National University of Singapore, 3 Science Drive 2, 117543 Singapore}
\affiliation{Temasek Laboratories, National University of Singapore, 5A Engineering Drive 1, 117411 Singapore}
\author{S. R. Chanu}
\affiliation{Centre for Quantum Technologies, National University of Singapore, 3 Science Drive 2, 117543 Singapore}
\author{R. Kaewuam}
\affiliation{Centre for Quantum Technologies, National University of Singapore, 3 Science Drive 2, 117543 Singapore}
\author{M. S. Safronova}
\affiliation{Department of Physics and Astronomy, University of Delaware, Newark, Delaware 19716, USA}
\affiliation{Joint Quantum Institute, National Institute of Standards and Technology and the University of Maryland,
College Park, Maryland, 20742}
\author{M. D. Barrett}
\email{phybmd@nus.edu.sg}
\affiliation{Centre for Quantum Technologies, National University of Singapore, 3 Science Drive 2, 117543 Singapore}
\affiliation{Department of Physics, National University of Singapore, 2 Science Drive 3, 117551 Singapore}
\begin{abstract}
Branching fractions for decays from the $P_{3/2}$ level in $^{138}$Ba$^+$ have been measured with a single laser-cooled ion.  Decay probabilities to $S_{1/2}$, $D_{3/2}$ and $D_{5/2}$ are determined to be $0.741716(71)$, $0.028031(23)$ and $0.230253(61)$, respectively, which are an order of magnitude improvement over previous results.  Our methodology only involves optical pumping and state detection, and is hence relatively free of systematic effects.   Measurements are carried out in two different ways to check for consistency.  Our analysis also includes a measurement of the $D_{5/2}$ lifetime, for which we obtain 30.14(40)\,s.
\end{abstract}
\maketitle
\section{Introduction}
Singly-ionized Barium has been well studied over the years with a wide range of precision measurements \cite{marx1998precise,knoll1996experimental,hoffman2013radio,woods2010dipole,sherman2008measurement,kurz2008measurement,munshi2015precision,dutta2016exacting} providing valuable benchmark comparisons with theory \cite{guet1991relativistic,dzuba2001calculations,iskrenova2008theoretical,safronova2010relativistic,sahoo2007theoretical,gopakumar2002electric}.  Such comparisons have been motivated in part by the proposed parity nonconservation (PNC) measurement using the $S_{1/2}-D_{3/2}$ transition in $^{137}$Ba$^+$ \cite{fortson1993possibility}.  Recently, we have proposed that accurately measured properties of $^{138}$Ba$^+$ can be combined to provide an accurate model of the dynamic differential scalar polarizability $\Delta \alpha_0(\omega)$ for the $S_{1/2}-D_{5/2}$ clock transition \cite{BarrettProposal}.  With such a model, ac-Stark shifts of the clock transition could provide an \emph{in situ} calibration of laser intensities and comparison of polarizabilities with another species.  A limiting factor for the proposal was the accuracy of existing values for $P_{3/2}$ branching ratios.  Moreover, recent measurements of $P_{1/2}$ branching fractions \cite{arnold2019measurements} indicated possible problems with existing $P_{3/2}$ values \cite{dutta2016exacting}.  Consequently, we have remeasured these values with an order of magnitude improvement in their uncertainties.

Our measurements only involve optical pumping and state detection.  Consequently, results are insensitive to laser intensity, frequency, and polarization changes provided they do not significantly affect optical pumping times and detection.  Our implementation also allows the branching ratios to be determined in two different ways, which we do by way of a consistency check on our results.
\section{Experiment Overview}
\begin{figure}
\begin{center}
  \includegraphics[width=\linewidth]{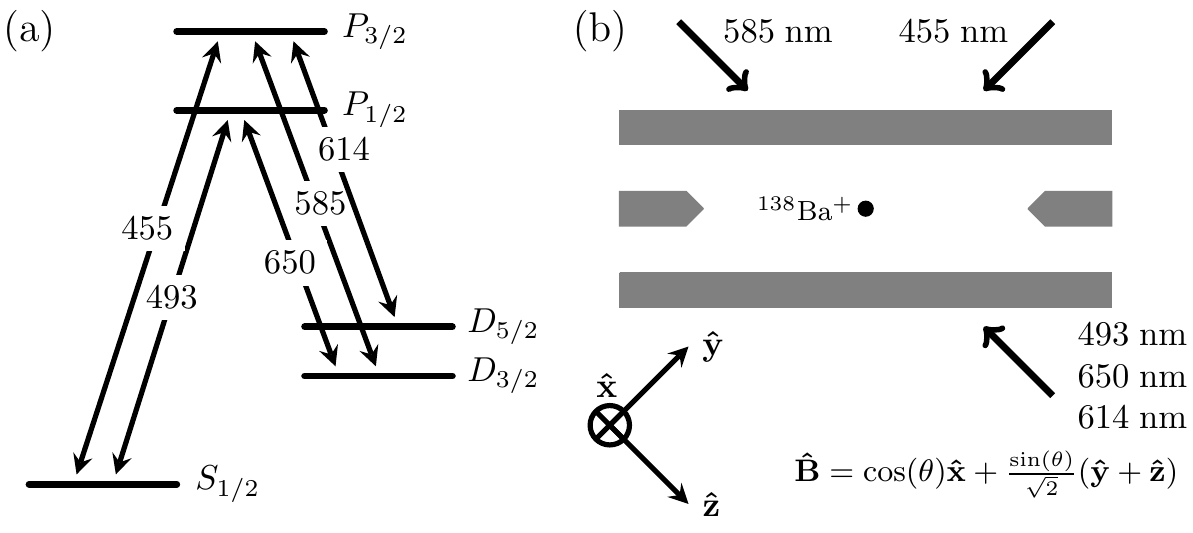}
  \caption{(a) Level structure showing the Doppler cooling and detection transitions at 493 and 650\,nm, and optical pumping transitions at 455, 585, and 614\,nm, (b) Beam configurations relative to the trap. The magnetic field is specified with respect to the axes shown and has $\theta\approx 33^\circ$.}
  \label{Fig1}
\end{center}
\end{figure}

The relevant level structure for Ba$^+$ is given in Fig.~\ref{Fig1}(a).  Doppler cooling is provided by driving the transitions at 493 and 650\,nm, with light scattered at 650\,nm collected onto a single photon counting module (SPCM) for detection.  The remaining three transitions at 455, 585, and 614\,nm facilitate optical pumping as required.  Throughout the report, we refer to $S_{1/2}$ and $D_{3/2}$ states collectively as the bright state, and $D_{5/2}$ states as the dark state.  

The experiments are carried out in a linear Paul trap with axial end-caps as described in \cite{arnold2018blackbody,kaewuam2018laser}.  The laser configurations relative to the trap are shown in Fig.~\ref{Fig1}(b).  An applied magnetic field of $\sim0.31\,\mathrm{mT}$ with the orientation given in Fig.~\ref{Fig1}(b) is sufficient to avoid any dark states, independent of the polarization state of any of the lasers.

\begin{figure*}
\begin{center}
  \includegraphics[width=1.0\linewidth]{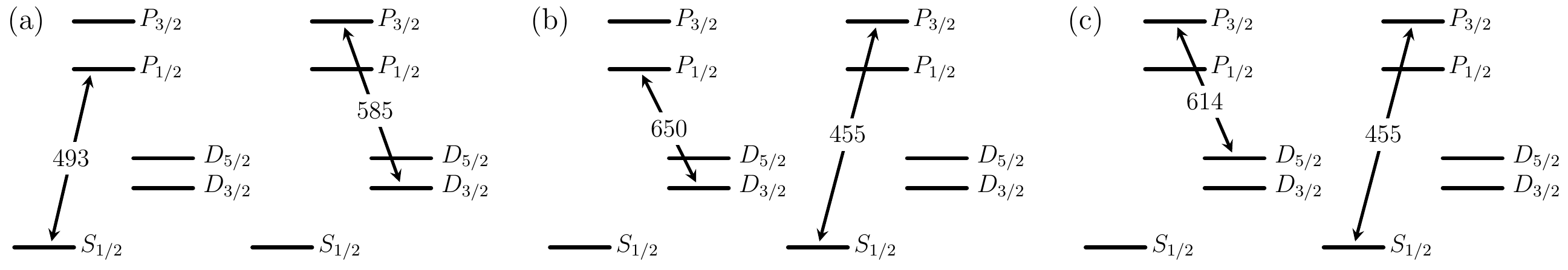}
  \caption{Optical pumping schemes used in this work. (a), (b), and (c) provide measurements of $p$, $q$, and $1-r$, respectively, with $p,q,r$ as defined in the text.  Population is given to be distributed in $S_{1/2}$ and $D_{3/2}$ at the start for (a) and (b), and optically pumped to $D_{5/2}$ for (c).  In each case the two steps given can be repeated with $m$ repetitions giving populations $p^m$, $q^m$, and $1-r^m$ in the bright states $S_{1/2}$ and $D_{3/2}$.}
\label{Pumping}
\end{center}
\end{figure*}

Probabilities for decay from $P_{3/2}$ to $S_{1/2}$, $D_{3/2}$, and $D_{5/2}$ are here denoted $p_1, p_2,$ and $p_3$, respectively.  They are inferred from population measurements following the optical pumping sequences shown in Fig.~\ref{Pumping} where it is given that each sequence is preceded by preparation into the appropriate bright or dark state. The remaining bright state population for the sequences shown in (a), (b), and (c), are $p$, $q$ and $1-r$, respectively, where 
\begin{gather}
p=\frac{p_1}{p_1+p_3}, \quad q=\frac{p_2}{p_2+p_3},\\
\intertext{and}
r=\frac{p_1 p_3}{(1-p_1)(1-p_3)}.
\end{gather}
Since $\sum p_k=1$, only two measurements are needed to uniquely determine $p_k$. In addition, each of the sequences in Fig.~\ref{Pumping} can be repeated $m$ times within a single experiment to provide measurements of $p^m$, $q^m$, and $1-r^m$.  This can provide better statistics with a similar approach having been demonstrated with Ca$^+$ \cite{gerritsma2008precision}.  Here we provide measurements of $p$ and $q$, and use measurements of $p^m$ and $1-r^m$ over a range of $m$ to check for consistency.
\section{Measurements}
\label{Measurements}
\subsection{Measurement of q}
\label{qMeasure}
Measurements of $q$ involved optical pumping with lasers at 650, 455, and 614\,nm with measured pumping time constants of 0.3, 1.0, and $1.3\,\mathrm{\mu s}$, respectively.  However, as the 455-nm laser was only locked by a wavemeter, occasional frequency excursions of $\sim 10\,\mathrm{MHz}$ could occur, significantly changing the pumping rate.  Consequently, the pump rate was also measured within each cycle of the experiment.  In addition, measurements of a dark and bright state were also made to assess the detection efficiency in each case.  Thus, a single experiment cycle consisted of four interleaved experiments.

Explicit pulse sequences for each of the four experiments are given in table~\ref{qExpt}.  The cooling pulse in all four cases had a duration of $400\,\mathrm{\mu s}$ and included repump light at 614\,nm to ensure the atom is pumped to the bright state.   The first two experiments use lasers at 455 and 650\,nm to optically pump to $D_{5/2}$; partial pumping with a $1\,\mathrm{\mu s}$ pulse monitors pumping rate of the 455-nm transition, and complete pumping with a $30\,\mathrm{\mu s}$ pulse prepares the ion in the dark state.  The third experiment uses complete optical pumping of the 455-nm transition alone to determine $q$, and the final experiment provides a measurement of the bright state.  The full cycle was repeated in blocks of 1000.  Between each block, 10 measurements of the counts in 1\,ms were made for both the bright state and background, where the latter was taken with the 493-nm laser turned off.
\begin{table}
 \caption{Interleaved experiments made when measuring $q$ and the pulse sequences used for each.  Cooling pulses had a duration of $400\,\mathrm{\mu s}$ and included repump light at 614\,nm in addition to the detection light at 493 and 650\,nm.}
 \label{qExpt}
\begin{ruledtabular}
\begin{tabular}{llll}
Expt. 1 &  Expt. 2 & Expt. 3 & Expt. 4 \\
 \vspace{-0.4cm}\\
 \hline
 \vspace{-0.4cm}\\
Cooling & Cooling & Cooling & Cooling \\
650 ($20\,\mathrm{\mu s}$) & 650 ($20\,\mathrm{\mu s}$) &650 ($20\,\mathrm{\mu s}$) &-\\
650/455 ($1\,\mathrm{\mu s}$) & 650/455 ($30\,\mathrm{\mu s}$) &455 ($30\,\mathrm{\mu s}$) &-\\
Detect. & Detect. & Detect. & Detect. \\
 \end{tabular}
 \end{ruledtabular}
 \end{table}

Over the course of two days, $\sim16300$ blocks of data were taken.  The data is conditioned on the 'bright detection' from expt~4, which gives a mean probability of $4.8\times10^{-5}$ for detecting dark.  These events are not, however, Poisson distributed.  In all cases where two or more dark events occured, they were clustered in consecutive experiments suggesting a collision event had inhibited bright detection through multiple 400\,$\mu$s cooling cycles. Consequently any blocks of 1000 cycles that have more than one dark state detection event during expt.~4 were omitted.  This eliminates only 15 blocks, or $\sim 0.1\%$ of the data.

\begin{figure}[ht]
\includegraphics[width= \linewidth]{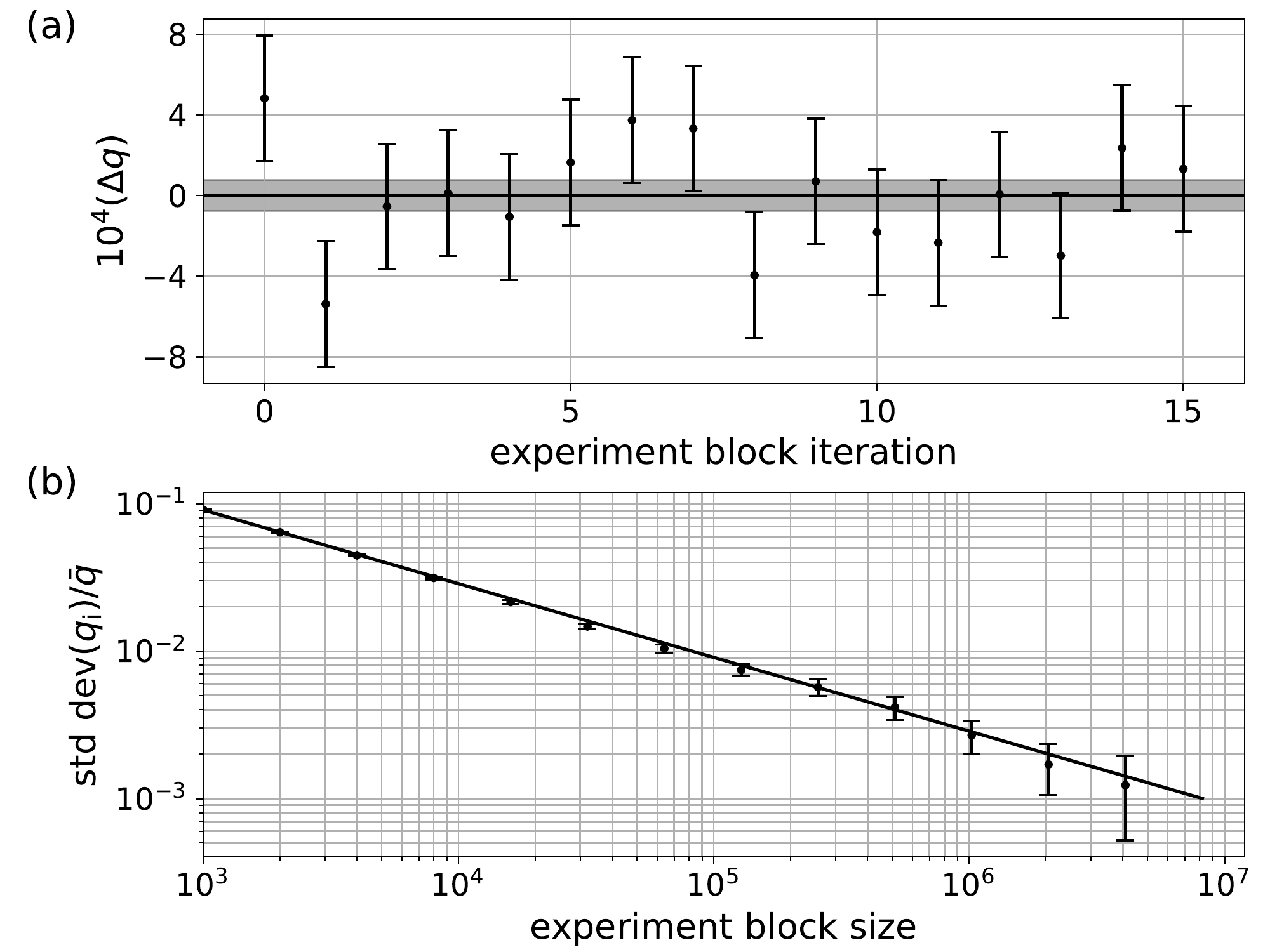}
\caption{Results for the measurements of $q$. (a) Mean values of $q$ estimated from blocks of size $N=10^6$, with $\Delta q$ denoting the value relative to the mean of all data.  Uncertainties for each point are the expected projection noise for the given $N$, and the shaded region is the uncertainty in the mean of all data. (b) Fractional standard deviation of mean values estimated from blocks of size $N$.}
\label{qData}
\end{figure}

From the remaining data we obtain an estimate $q=0.108527(77)$, where the uncertainty is statistical.  Mean values of $q$ estimated from blocks of size $10^6$ are given in Fig.~\ref{qData}(a) where the uncertainties for each point are the expected projection noise, and the reduced $\chi^2$ is 0.84. The fractional standard deviation of mean values estimated from blocks of variable size $N$ is given in Fig.~\ref{pData}(b) to further demonstrate that the measurement continues to average down in accordance with the projection noise limit given by the solid line.
\subsection{Measurement of p}
\label{pMeasure}
As illustrated in Fig.\,\ref{Pumping}\,(a), measurement of $p$ entails optical pumping with lasers at 493 and 585\,nm, which had measured pumping time constants of approximately 500 and 300\,ns, respectively.  The 585-nm light was generated by an 1170-nm diode laser that was locked to a frequency comb \cite{kaewuam2019spectroscopy} and frequency doubled by a fiber-coupled, waveguide doubler.  The result was a very stable scattering rate compared to that with the  455-nm laser used in the measurement of $q$.  Thus, the typical experiment cycle consisted of only two interleaved experiments: a bright state detection, to monitor collision events, and the measurement of $p$.  The latter consisted of $400\,\mathrm{\mu s}$ of Doppler cooling, $40\,\mathrm{\mu s}$ of optical pumping of the 493-nm transition to $D_{3/2}$, $40\,\mathrm{\mu s}$ of optical pumping with 585-nm light, and then detection.  The pumping rate of the 585-nm transition was additionally monitored in-between each  block to confirm its stability.  

\begin{figure}
\includegraphics[width= \linewidth]{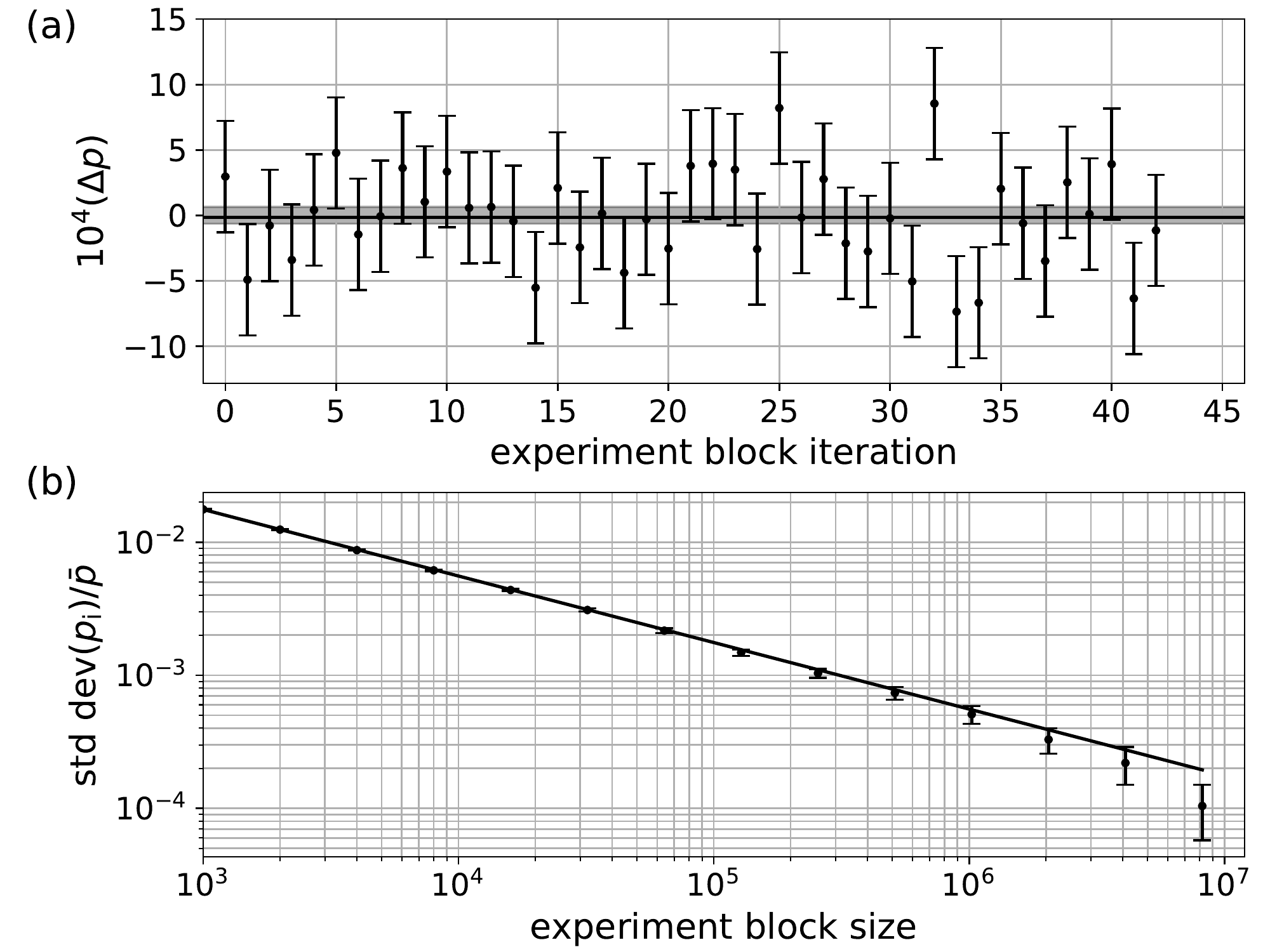}
\caption{Results for the measurements of $p$. (a) Mean values of $p$ estimated from blocks of size $N=10^6$, with $\Delta p$ denoting the value relative to the mean of all data.  Uncertainties for each point are the expected projection noise for the given $N$, and the shaded region is the uncertainty in the mean of all data. (b) Fractional standard deviation of mean values estimated from blocks of size $N$.}
\label{pData}
\end{figure}

As for the measurement of $q$, data is conditioned on the bright state detection experiment removing 16 out of 43100 blocks.  From the remaining $\sim43\,\mathrm{M}$ experiments, we obtain the estimate $p=0.763107(65)$.  Mean values of $p$ estimated from blocks of size $10^6$ are given in Fig.\,\ref{pData}\,(a) where the uncertainties are the expected projection noise, and the associated reduced $\chi^2$ is 0.76.  The fractional standard deviation of mean values estimated from blocks of variable size $N$ is given in Fig.~\ref{pData}\,(b) to further demonstrate the measurement continues to average down in accordance with the projection noise limit given by the solid line.
\subsection{Measurements of p\textsuperscript{m} and 1-r\textsuperscript{m}}
\label{Consistency}
\begin{figure}[ht]
\includegraphics[width= \linewidth]{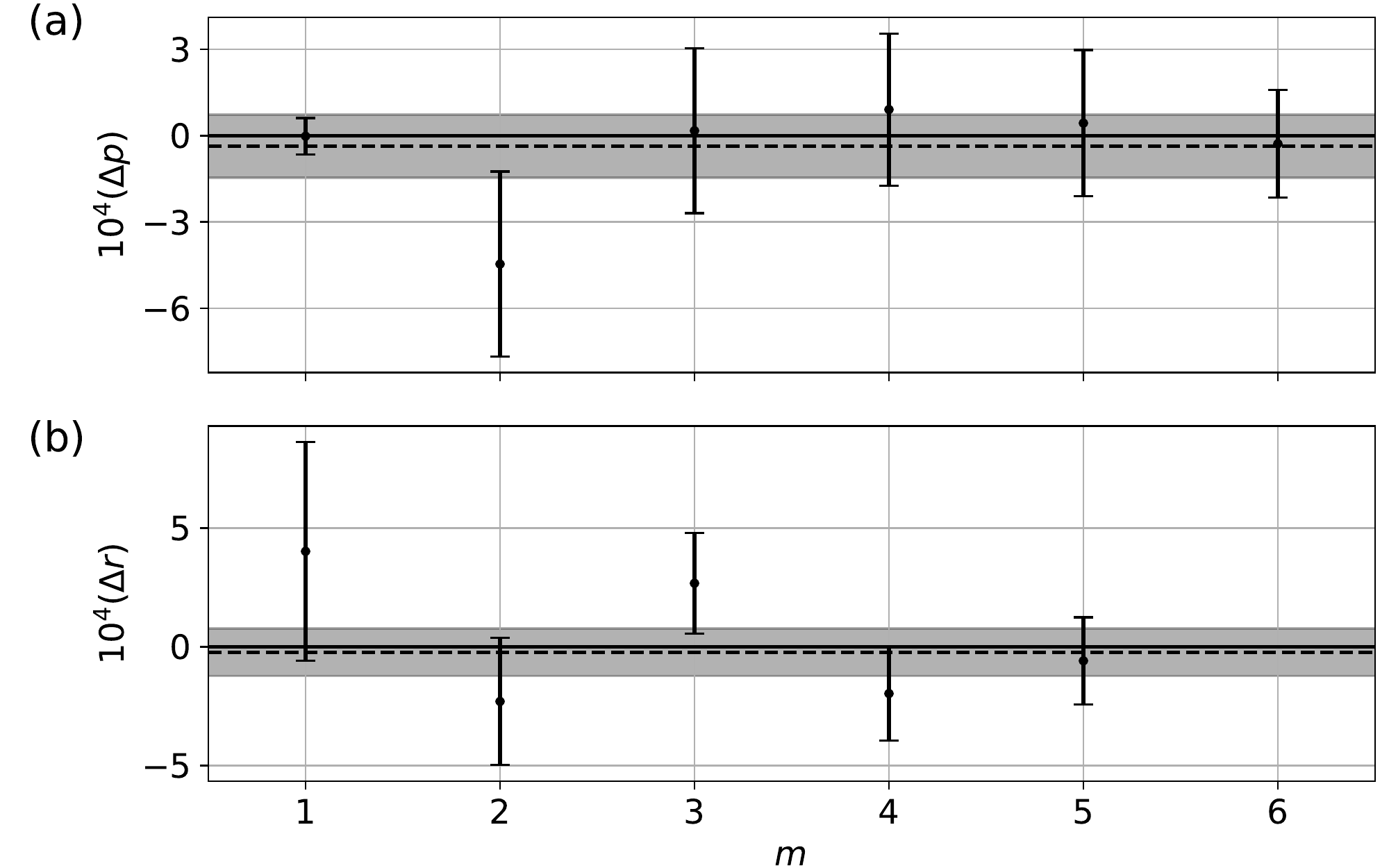}
\caption{Estimated values of $p$ and $r$ from measurements of $p^m$ and $1-r^m$ where $m$ is the number of pumping cycles within a single experiment as discussed in the text. Values for $p$ are relative to the $m=1$ result found in Sect.~\ref{pMeasure}.  Values for $r$ are given relative to the value found using Eq.~\ref{rEquation} and values of $p$ and $q$ from Sect.~\ref{qMeasure}, and~\ref{pMeasure}.  In both cases, the dashed line is the weighted mean over $m$ and the shaded region the corresponding uncertainty. The weighted mean for $p$ excludes the $m=1$ result.}
\label{mCheck}
\end{figure}

In a similar manner to the previous two subsections, consistency checks were carried out by cycling the pumping schemes in Fig.~\ref{Pumping}\,(a) and (c) to measure $p^m$ and $1-r^m$ for a range of $m$.  Values for the estimated $p$ and $r$ for each value of $m$ are plotted in Fig.~\ref{mCheck} where each measurement consists of approximately $10^6$ experiments.  Taking the weighted average over $m$ in each case gives $p=0.76307(11)$ and $r=0.85899(10)$, where we have excluded the $m=1$ result in the estimate for $p$.

The value for $p$ is in good agreement with $m=1$ result $p=0.763107(65)$ given in the previous section.  The value of $r$ is also in good agreement with $r=0.85901(10)$ obtained from the expression
\begin{equation}
\label{rEquation}
r=\frac{p(1-q)^2}{p+q-2p q},
\end{equation}
and the measured values of $p$ and $q$ from the previous sections.
\subsection{The D\textsubscript{5/2} lifetime}
\label{Lifetime}
A recent measurement \cite{dijck2018lifetime} gave a $D_{5/2}$ lifetime of 25.6(0.5)\,s, which was significantly less than previous experimental reports \cite{auchter2014measurement,gurell2007laser,royen2007monitoring,nagourney1986shelved,madej1990quantum} and numerous calculations \cite{safronova2017forbidden}.  As it factors into consideration of systematic effects, we provide our own measurement for reference purposes.  Our method continuously monitored a string of four ions initialized with typically one or two ions in the bright state and registered when each ion became bright.  Detection light was imaged onto an electron multiplying charged-coupled device (EMCCD) camera with an imaging system having sufficient resolution to resolve the state of individual ions from a 10\,ms exposure with typically better than $99\%$ efficiency.  Images were captured every 20\,ms until all ions were detected bright, which was verified by a further ten images.  In these experiments, the 614 light was blocked at all times.

The initial state of the ion-string was set by optical pumping to $\ket{S_{1/2},m=1/2}$ and driving $\ket{S_{1/2},m=1/2}\leftrightarrow\ket{D_{5/2},m=1/2}$ with a clock laser at 1762\,nm.  Ideally, the initial string would have one bright ion to facilitate sympathetic cooling for other ions in $D_{5/2}$.  However, the configuration could only be done probabilistically, which depends on the spatial profile of the beam, and pulse duration of the clock laser.  Instances in which no ions are transferred to $D_{5/2}$ simply trigger a restart.  Records starting with all ions dark are still used, so long as ions appear bright one at a time and the crystal is stable.

Within the detection records, there are four identifiable events in addition to the expected transition to the bright state: detection errors, swaps, melts, and dark transitions.  Detection errors are identified by a single image that is different from previous and subsequent images.  Swaps, as the name would imply, are a change in the ion configuration without a change in number of bright ions.  Melts are characterized by all ions appearing dark for several images.  They are presumably the result of a high impact collision, which destabilizes the ion crystal and may take several seconds to recrystallize.  Finally, dark transitions are identified by a decrease in the number of bright ions and are the result of very rare, off-resonant excitation of the 493-nm cooling light.  

Approximately 32 hours of data were collected from the same string of four ions with a total 2429 detection records starting with at least one dark ion.  An additional 48 records were discarded as quantum jumps could not be unambiguously identified due to: (a) multiple ions appearing bright when starting from an all dark state (9); (b) the all bright verification step failing indicating a detection error triggered the last event (25); (c) the number of bright ions changing after a melting event (11); or (d) multiple ions transitioning at the same time (3).  

\begin{figure}
\includegraphics[width=\linewidth]{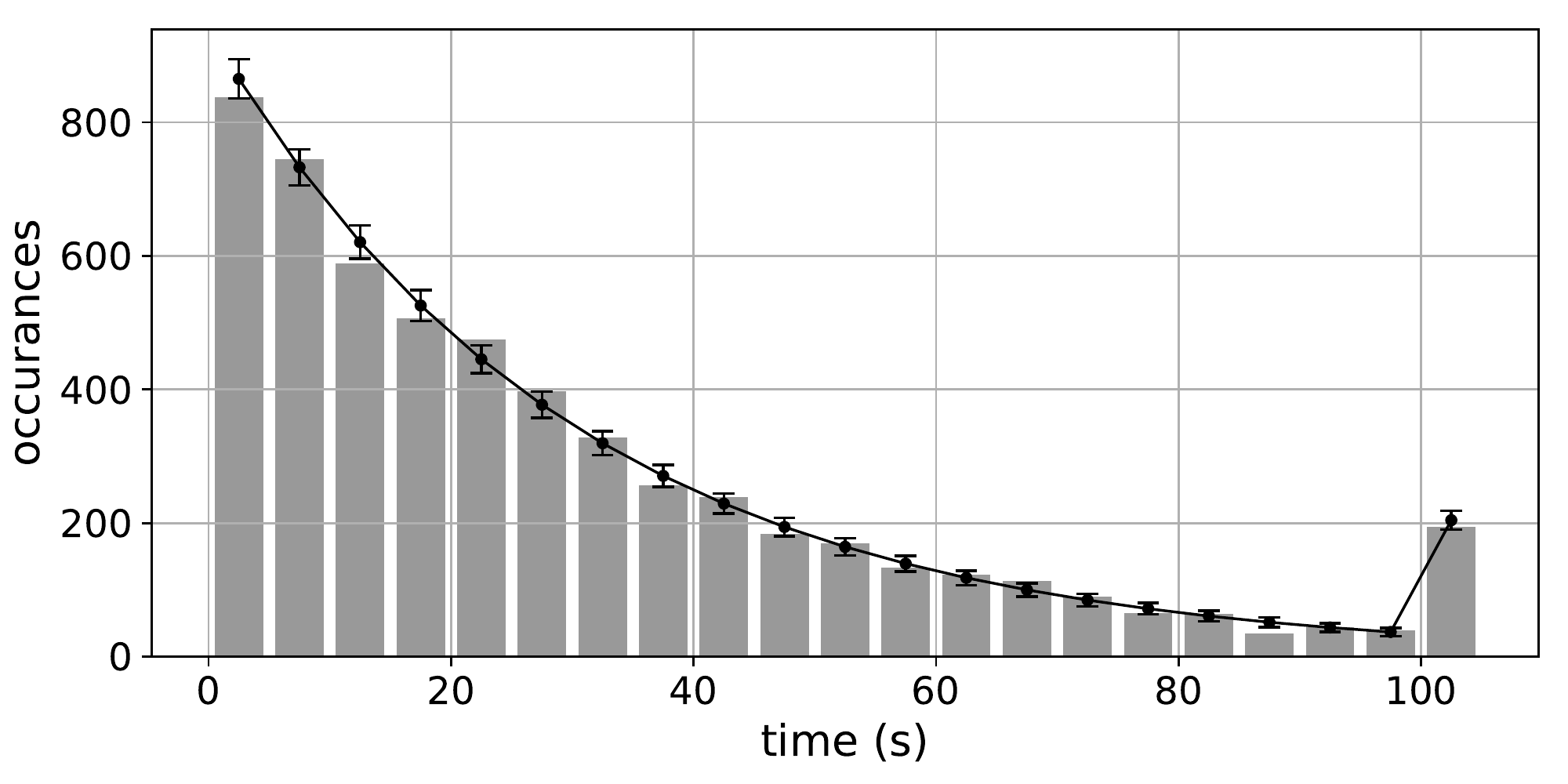}
\caption{Histogram of observed decay times from $D_{5/2}$.  Black dots are the expected values from the estimated $\tau_D=30.14(40)$ and number of events recorded, with the uncertainty indicating the expected standard deviation. The reduced $\chi^2$ for the observed histogram is 0.99.}
\label{LifetimeData}
\end{figure}

Within the 2429 detection records there were 5620 bright transitions, 179 detection errors, 253 swaps, 50 melts, and 20 dark transitions.  As all quantum jumps are recorded, the average of all bright transition times, $\tau_D=30.14(40)\,\mathrm{s}$, is the unbiased estimate of the lifetime.  In Fig.~\ref{LifetimeData}, we give a histogram of all recorded bright transition times in 5\,s bins.  The final bin is the total of all events over 100\,s.  Black dots and the uncertainties are theoretical expectations given the estimated $\tau_D$ and number of events recorded.  The reduced $\chi^2$ for the observed histogram is 0.97.

Note that it is important that all quantum jumps are observed in a given record when using this approach.  When monitoring the decay of $N$ ions, the time for the first decay is governed by an exponential distribution with parameter $N\lambda$, where $\lambda$ is the decay rate for one ion.  Hence, an observed set of times $\{T_k\}$ has a distribution
\begin{align}
P(\{T_k\})&=e^{-N \lambda T_1}e^{-(N-1) \lambda (T_2-T_1)}\cdots e^{-\lambda (T_N-T_{N-1})}\nonumber\\
&=\prod_{k} e^{-\lambda T_k},
\end{align}
which is the expected $N$ independent, identical distributions for a single decay.  This would not be the case if some events were missed.

With the exception of the most recent report \cite{dijck2018lifetime}, the value of $\tau_D$ is consistent with previous reports \cite{auchter2014measurement,gurell2007laser,royen2007monitoring,nagourney1986shelved,madej1990quantum} and a number of theoretical estimates \cite{safronova2017forbidden}.  Collision-induced decays are not likely significant in this experiment; of the 5620 observed decays, only 12 included an observable change in the ion configuration.  In any case, such effects would decrease the observed lifetime.  
\section{Systematics}
The experiments only involve optical pumping and state detection, which results in relatively few considerations for systematic effects.  These are finite optical pumping times, unwanted scattering from leakage light, finite lifetime of $D_{3/2}$ and $D_{5/2}$, collisions, and detection errors.  The effect of finite lifetimes for $D_{3/2}$ and $D_{5/2}$ have the same effect as leakage light at 650 and 614\,nm, respectively, and are considered in this context.  Similarly collisions and detection errors  are also considered in a similar context.
\subsection{Finite pumping times}
With the exception of the lasers at 455 and 614\,nm, which were locked to a wavemeter, all others lasers were locked to stable references giving correspondingly stable pumping rates. In the latter cases pumping times were more than 20 time-constants of the associated pumping rates, corresponding to a probability of $<2 \times 10^{-9}$ for pumping errors.  Hence only the 455- and 614-nm pumping times might be of concern.

\begin{figure}
\includegraphics[width=0.9 \linewidth]{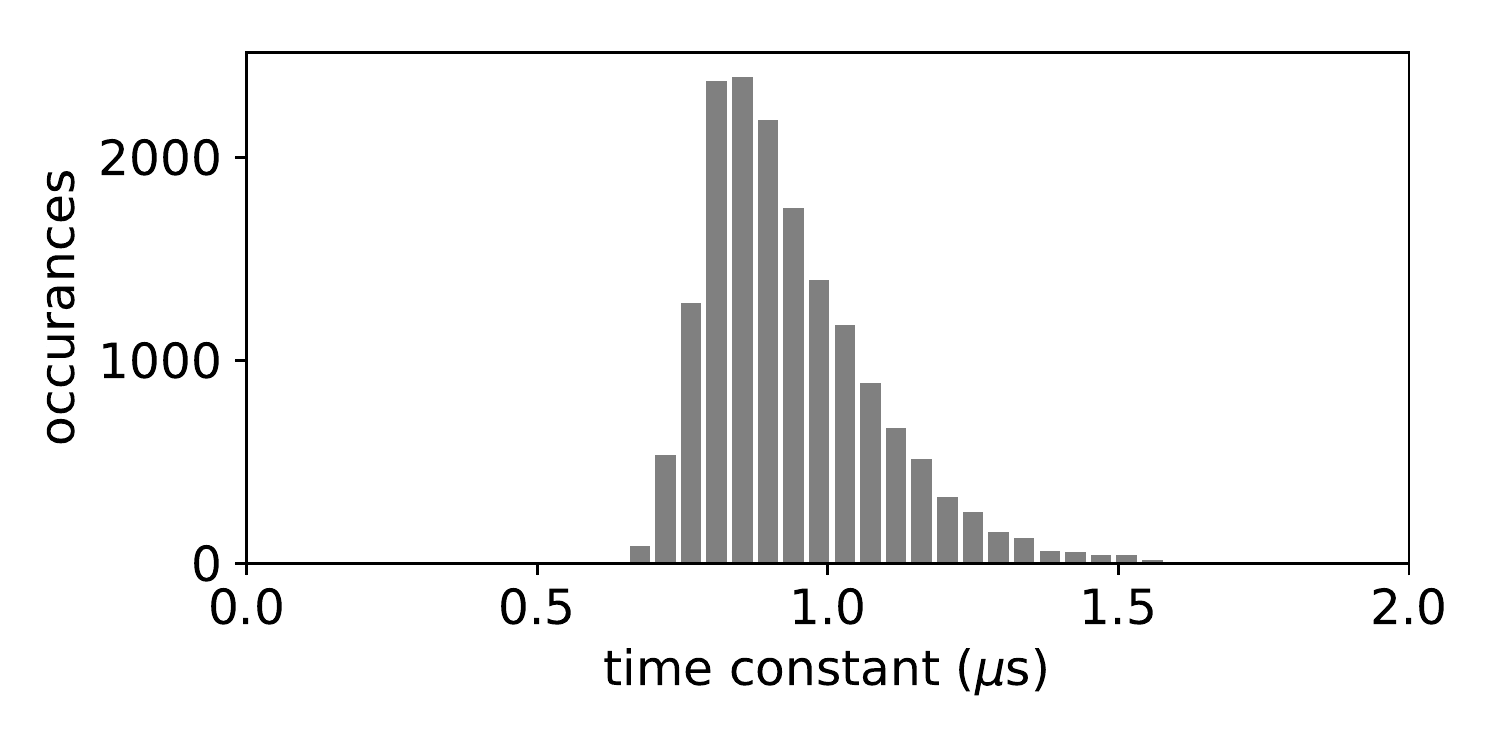}
\caption{Histogram of the estimated time constants associated with the 455-nm pumping rate determined throughout the measurement of $p$.  These can be compared to the total pumping time of $40\,\mathrm{\mu s}$ used in the experiment.}
\label{Rate455}
\end{figure}

As noted in Sect.~\ref{pMeasure}, the 455-nm pumping rate was monitored in an interleaved experiment during the measurement of $p$ by measuring the population after partial pumping for a fixed time.  From these measurements the pumping rate was estimated and a histogram of the associated time constant is given in Fig.~\ref{Rate455}.  Almost all experiment cycles have pumping times more than 20 time-constants of the measured pumping rates rendering this effect is negligible at the $\lesssim 10^{-8}$ level.  
Repumping on the 614-nm transition occurs during the 400 $\mu$s cooling which is far longer than the pumping time constant and thus insensitive to variations.  The pumping rates for transitions at 455 and 614\,nm were not monitored during the measurements of $p^m$ and $1-r^m$ in Sect.~\ref{Consistency}.  However the consistency of those results with the independently measured $p$ and $q$ demonstrate that the finite pumping times are not important to the accuracy reported.
\subsection{Leakage pumping rates}
Here we describe the effects of stray light from each of the lasers and the experiments used to bound their effects.  If a laser was not needed for any measurements in Sect.~\ref{Measurements}, it is given that it was physically blocked and not merely switched off via an acousto-optic modulator (AOM).
 
\subsubsection{Stray light at 455\,nm}
The 455-nm laser was derived from a 910-nm diode laser, which was switched by a double-pass acousto-optic modulator (AOM) before a single-pass frequency doubling crystal.  Hence the extinction ratio was very high. This was tested by optical pumping into $S_{1/2}$, and waiting 10\,ms before detection and cooling (with light at 614\,nm blocked).  In $\sim672\,\mathrm{k}$ cycles we observed one event pumping to $D_{5/2}$ with three others attributed to collisions.  The latter are identified by a sudden drop in counts that slowly recovers to the normal level. 

The main effect of stray light at 455\,nm would be to influence bright state detection, by pumping into the dark state, and measurements made of $r^m$, by repumping during pumping of the 614-nm transition.  In any case,  given a pumping rate of $\lesssim 1/\mathrm{hour}$ and the typical experiment time of $<$400 $\mu$s including pumping and detection, the probability of unwanted pumping is $\lesssim 1\times 10^{-7}$.

\subsubsection{Stray light at 614\,nm}
Stray light from the 614-nm laser effectively decreases the lifetime of the $D_{5/2}$ level, which was measured to be 30.14(40)\,s on a four ion string, as described in Sect.~\ref{Lifetime}.  To test the leakage rate, we repeated this lifetime measurement with the 614-nm light unblocked but the switching AOM off.  A neutral density filter attenuating the 614-nm beam was removed to give maximum pumping rate with the AOM on, which was measured to be $\sim$420\,ns at the time of this test.  From 239 decay events, we get an estimated lifetime of 29.7(1.9)\,s.  This implies a scattering rate of $0.7 \pm 2.2 \times10^{-3}\,\mathrm{s}^{-1}$ consistent with zero. Taking $3\times10^{-3}\,\mathrm{s}^{-1}$ as an estimated upper bound, the probability of scattering during the experiment is $\lesssim 1 \times 10^{-6}$. During the actual data runs the measured pumping time constant was $\sim 1\,\mathrm{\mu s}$ so the leakage scattering rate likewise would be smaller. 

\subsubsection{Stray light at 585\,nm}
Similar to the 455-nm laser, the switch AOM for the 585-nm beam was placed before the doubling stage resulting in high extinction.  As an initial test, the ion was left continuously scattering with the 585-nm light `off', but unblocked, for about half an hour with no dark events.  A further test was done by pumping to  $D_{3/2}$ and waiting for\,10 ms before detection and cooling (with light at 614\,nm blocked).  In 480\,k experiments, we observed only one pumping event and two collisions.  As with stray light at 455\,nm, given a pumping rate of $\lesssim 10^{-3}\,\mathrm{s}^{-1}$ and typical experiment time of $<$400 $\mu$s including pumping and detection, the probability of unwanted pumping is then $\lesssim 4 \times 10^{-7}$. 

\subsubsection{Stray light at 650\,nm}
Stray light from the 650-nm laser effectively decreases the lifetime of of $D_{3/2}$.  Unlike light at 455, 614, and 585\,nm, stray light at 650\,nm does not affect results at the detection step and thus one is only concerned with scattering during the pumping steps.  A test for any leakage light at 650\,nm was done by pumping to $D_{3/2}$ and waiting for $\tau_w=20\,\mathrm{ms}$ with light at 455\,nm on before detection and cooling.  If a scattering event on the 650-nm transition results in a decay into $S_{1/2}$, it is then pumped to $D_{5/2}$ by light at 455\,nm with probability $(1-q)\sim0.9$.

From 87.2\,k cycles, we measured a dark probability of $p_d=2.85(18)\times10^{-3}$.  From the 79.8(4.6)\,s lifetime of $D_{3/2}$, we would expect a 2.5(3)$\times10^{-4}$ probability of decay to $S_{1/2}$ in the 20 ms, of which $1-q$ will be pumped dark by light at 455\,nm. The observed rate is thus dominated by scattering on the 650-nm transition, and the rate of scattering out of $D_{3/2}$ is $\gamma_s=p_d/(\tau_w (1-q))=0.15\,\mathrm{s}^{-1}$.  This effectively reduces the lifetime of $D_{3/2}$ to about $6.5\,\mathrm{s}$.

For measurements of $p$, there is little opportunity for leakage light at 650\,nm to cause a problem because light at 585\,nm will depopulate $D_{3/2}$ on a timescale of $1 \mu$s.  So the error caused by leakage light at 650\,nm can only be $\sim 1 \times 10^{-7}$.  During measurements of $q$, decay of the population from $D_{3/2}$ during the $30\,\mathrm{\mu s}$ of optical pumping on the 455-nm transition will be more significant.  Since the 455-nm optical pumping rate is much shorter than the pumping time, we make the approximation that any $D_{3/2}$ decay is immediately pumped to $D_{5/2}$ with probability $1-q$.  The error is then $\gamma_s \times 30\,\mu\mathrm{s}\times (1-q) \approx 4 \times 10^{-6}$.

\subsubsection{Stray light at 493\,nm}
The 493-nm beam is double-passed by two AOMs and the measured extinction was in excess of 130 dB.  We nonetheless attempted to measure this by pumping to $S_{1/2}$, waiting 10 ms before optical pumping on the 585-nm transition before detection and cooling, all with light at 614\,nm blocked.  From 40\,k cycles we saw no shelving events.   Thus the scattering rate out of $S_{1/2}$ would be $\lesssim10^{-3}\,\mathrm{s}^{-1}$.  As with stray light at 650\,nm, this does not affect detection, and hence the error is $\lesssim 4\times 10^{-8}$.
\subsection{Collisions and detection errors}
For the branching ratio measurements reported here we used a Bayesian detection scheme \cite{paez2016atomic,myerson2008high} in which the bright (dark) state probability is estimated in real time and terminates when the bright (or dark) state probability falls below a preset threshold, here set to $10^{-6}$. Count rates during the experiments were typically in the range $\sim35-40$ counts/ms for a bright state, and a more stable 0.4 counts/ms for a dark state.

From the measurements sequences used for $q$, the observed probability of detecting dark during expt.~4 was $7.2(7)\times10^{-6}$, which is taken as an estimate of a bright state detection error. Similarly, the probability of detecting a bright state during expt.~2 was $1.9(1)\times10^{-5}$, which is taken as an estimate of a dark state detection error. In both cases the observed error rates are significantly above the $10^{-6}$ thresholds set in the Bayesian detection algorithm.  Possible causes for excess detection errors would be collisions and the $D_{5/2}$ lifetime for bright and dark states, respectively.

The bright state control experiments taken during the measurements of $p$, and $q$ and records for the $D_{5/2}$ lifetime measurement gave clear evidence of collisions.  From the collision events identified in the measurements of $p$ and $q$, the estimated collision rate would be roughly 1 every 48\,minutes.  Similarly data from the $D_{5/2}$ lifetime suggests a collision rate of 1 every 6\,minutes, which is roughly consistent given the larger number of ions.  Collision events occurring on this timescale, however, would not significantly contribute to the observed error rates, more so that these events were detected and removed.  However, those events were collisions with sufficient impact to interrupt cooling over many experimental cycles.  Less impactful collisions that interrupt only a single experiment would likely have a higher rate.  We also note that varying fluorescent rates can influence the validity of the threshold.  Simulations and independent experiments indicate this can be as much as a factor of four for a 10\%  drop in fluorescence.  In any case, the observed bright state error rate is not significant at the level of the statistical uncertainties in the reported results.

For the dark state control experiment, decay during detection would likely trigger a bright state result.  The probability of this occurring depends on the distribution of detection times. For our detection parameters, the average time to detect a dark state is $\sim 400\,\mu$s and the expected probability to decay within this time is 1.3$\times10^{-5}$.  Although clearly a contributing factor, it is still smaller than the observed error rate.  This could be due to state preparation errors associated with pumping into the dark state but, in any case, the observed dark state error rate of $<2\times 10^{-5}$ is not significant at the level of the statistical uncertainties in the reported results.  
\section{Discussion}
The dominant systematic in the branching ratios reported here are from detection errors and are not significant at the level of the statistical uncertainties.  Thus our final values are given by
\begin{equation}
\label{final}
p=0.763107(65), \quad \mbox{and}\quad q=0.108527(77).
\end{equation}
From these we can calculate the probabilities for decay from $P_{3/2}$ to $S_{1/2}$, $D_{3/2}$, and $D_{5/2}$, denoted $p_1, p_2,$ and $p_3$, respectively.  Values are given in table~\ref{branching} along with those calculated from theory for comparison.  As discussed in \cite{arnold2019measurements}, theory values were calculated using a linearized coupled-cluster approach including single-double excitations with uncertainties estimated from additional calculation methods.
\begin{table}
\caption{Probabilities for decay from $P_{3/2}$ to $S_{1/2}$, $D_{3/2}$, and $D_{5/2}$, which are denoted $p_1, p_2,$ and $p_3$, respectively.  Values are determined from the measured values of $p$ and $q$ given in Eq.~\ref{final}.  Theory values were estimated as discussed in \cite{arnold2019measurements}.}
 \label{branching}
\begin{ruledtabular}
\begin{tabular}{lccc}
\vspace{0.05cm}
 &  $p_1$ & $p_2$ & $p_3$ \\
 \hline
Expt.  & 0.741717(71) & 0.028031(23) & 0.230253(62) \\
Theory & 0.7423(18) & 0.0280(2) & 0.2297(15)\\
 \end{tabular}
 \end{ruledtabular}
 \end{table}

The experimental values of $p_k$ differ by 2-3 standard deviations of the results reported in \cite{dutta2016exacting}, but are in reasonable agreement with the values predicted in our previous work \cite{arnold2019measurements}.  The latter used measured branching fractions for decays from $P_{1/2}$, experimental matrix elements from \cite{woods2010dipole}, and theoretical estimates of matrix element ratios.  With the above branching fractions and the reduced matrix element $\langle P_{3/2}\|r\|S_{1/2}\rangle=4.7017(27)\,\mathrm{a.u.}$ reported in \cite{woods2010dipole}, matrix elements associated with $P_{3/2}$ decays and the lifetime can now be obtained using only experimentally determined values.  The remaining matrix elements are found to be
\begin{subequations}
\label{MatrixElements}
\begin{align}
\langle P_{3/2}\|r\|D_{5/2}\rangle&=4.1028(25)\,\mathrm{a.u.},\\
\langle P_{3/2}\|r\|D_{3/2}\rangle&=1.33199(96)\,\mathrm{a.u.},
\end{align}
\end{subequations}
which yields a lifetime of $\tau(D_{5/2})=6.2615(72)\,\mathrm{ns}$ or linewidth $\Gamma=2\pi\times 25.418(29)\,\mathrm{MHz}$.  For all quantities, the dominant contribution to the uncertainties is from the reduced matrix element,  $\langle P_{3/2}\|r\|S_{1/2}\rangle$.  

The matrix elements given in Eq.~\ref{MatrixElements} are in excellent agreement with theoretical calculations \cite{BarrettProposal}, but differ from estimates obtained using experimental results combined with ratios of matrix elements calculated from theory \cite{arnold2019measurements}.  The ratios in question are
\begin{subequations}
\label{Rk}
\begin{align}
R_0&=\frac{\langle{6 P_{3/2}}\|r\|{5 S_{1/2}}\rangle}{\langle{6 P_{1/2}}\|r\|{5 S_{1/2}}\rangle},\\
R_1&=\frac{\langle{6 P_{3/2}}\|r\|{5 D_{5/2}}\rangle}{\langle{6 P_{1/2}}\|r\|{5 D_{3/2}}\rangle},\\
\intertext{and}
R_2&=\frac{\langle{6 P_{3/2}}\|r\|{5 D_{3/2}}\rangle}{\langle{6 P_{1/2}}\|r\|{5 D_{3/2}}\rangle},
\end{align}
\end{subequations}
which can now be obtained from the experimental results in Ref.~\cite{woods2010dipole}, the $P_{1/2}$ branching ratio reported in Ref.~\cite{arnold2019measurements}, and values given in Eq.~\ref{final}.  Specifically we have
\begin{gather}
R_1=\sqrt{\left(\frac{1-p}{p}\right)\left(\frac{1-\bar{p}}{\bar{p}}\right)}\left(\frac{\omega_{455}\omega_{650}}{\omega_{614}\omega_{493}}\right)^{3/2}R_0,\\
\intertext{and}
R_2=\sqrt{\frac{q}{1-q}}\left(\frac{\omega_{614}}{\omega_{585}}\right)^{3/2}R_1,
\end{gather}
where $\bar{p}$ is the branching fraction for the $P_{1/2}$ level reported in \cite{arnold2019measurements}.  

Values for $R_0$, $R_2/R_1$, $R_1/R_0$, and $R_2/R_0$ obtained from experimental results are given in table~\ref{ratios}.  Uncertainties are derived in the usual manner by taking the quadrature sum of independent contributions with the exception of $R_0$ for which relative uncertainties from each matrix element are added to allow for possible correlation in their determination.  Theory values are also given for comparison with uncertainties determined from the maximum discrepancy between different computation methods \cite{arnold2019measurements}.  As noted in \cite{arnold2019measurements}, calculated ratios are expected to be accurate as they depend only weakly on correlation corrections.  Consequently, the $1.8\sigma$ and $1.6\sigma$ in the difference for $R_0$ and $R_1/R_0$ between experiment and theory maybe statistically significant.  We would therefore recommend a 2$\sigma$ estimate of uncertainties for derived quantities such as the matrix elements in Eq.~\ref{MatrixElements} and the corresponding estimates of $\tau$ and $\Gamma$.
\begin{table}
\caption{Matrix elements ratios $R_0$, $R_2/R_1$, $R_1/R_0$, and $R_2/R_0$, where $R_k$ are as given in Eq.~\ref{Rk}.  Experimental values are derived from the measured values of $p$ and $q$ given in Eq.~\ref{final}, matrix elements reported in \cite{woods2010dipole}, and the $P_{1/2}$ branching fraction reported in \cite{arnold2019measurements}.  Theory values were estimated as discussed in \cite{arnold2019measurements}.}
\label{ratios}
\begin{ruledtabular}
\begin{tabular}{ccc}
\vspace{0.05cm}
 &  Expt & Theory \\
 \hline
 $R_0$ & 1.4140(17) & 1.4109(2)\\
$R_2/R_1$ & 0.32466(13) & 0.32473(32) \\
$R_1/R_0$ & 0.95406(22) & 0.95338(38) \\
$R_2/R_0$ & 0.30974(14) & 0.30959(18) \\
 \end{tabular}
 \end{ruledtabular}
 \end{table}

In general, theoretical calculations for Ba$^+$ are in excellent agreement with experimental results derived from the matrix elements reported in \cite{woods2010dipole} and branching ratios reported here and in \cite{arnold2019measurements}.  In all cases the matrix elements are better than $1\%$ of the experimentally determined values. As branching fractions provide ratios of matrix elements, all values are tied to the values of $\langle P_{3/2}\|r\|S_{1/2}\rangle$ and $\langle P_{1/2}\|r\|S_{1/2}\rangle$ reported in \cite{woods2010dipole}.  Thus, it would be of interest to confirm those results by a different methodology such as that demonstrated in \cite{hettrich2015measurement, arnold2019dynamic}.  However, it would likely be difficult to reach a comparable level of accuracy via that approach.

In summary we have provided new branching ratio measurements for decays from $P_{3/2}$.  The new values are more than an order of magnitude more accurate than those given in previous reports and measurements have been done in two different ways to check for consistency.  Together with results in \cite{woods2010dipole} and \cite{arnold2019measurements}, they provide a complete set of experimentally determined matrix elements for all the dominant contributions to the differential scalar polarizability, $\Delta \alpha_0(\omega)$ of the $S_{1/2}\leftrightarrow D_{5/2}$ clock transition.  In addition, we have provided a new measurement of the $D_{5/2}$ lifetime, which also improves upon previous values.

\begin{acknowledgements}
This work is supported by the National Research Foundation, Prime Ministers Office, Singapore and the Ministry of Education, Singapore under the Research Centres of Excellence programme. M. S. Safronova acknowledges the sponsorship of the Office of Naval Research, USA, Grant No. N00014-17-1-2252.
\end{acknowledgements}
\bibliography{BranchingP}
\bibliographystyle{unsrt}

\end{document}